\documentclass[a4paper,12pt]{article}
\usepackage[a4paper, margin=1in]{geometry}
\usepackage{amsmath}
\usepackage{graphicx}
\usepackage{hyperref}
\usepackage{float}
\usepackage{booktabs}
\usepackage{tabularx}
\usepackage{longtable}
\usepackage[version=4]{mhchem} %for usage of \ce
\newcommand{\FTH}{\textit{\textbf{First 30-Vol\% layer}}}
\usepackage{authblk}

\title{Machine learning descriptors for predicting the high temperature oxidation of refractory complex concentrated alloys}

\author[1,*]{Akhil Bejjipurapu}
\author[1]{Alejandro Strachan}
\author[1]{Kenneth H. Sandhage}
\author[1]{Michael S. Titus}

\affil[1]{School of Materials Engineering, Purdue University, West Lafayette, IN 47907, USA}

\affil[*]{Corresponding author: abejjipu@purdue.edu}

\begin{document}

\maketitle

\begin{abstract}
Refractory Complex Concentrated Alloys (RCCAs) can exhibit exceptional high-temperature strength, making such alloys promising candidates for high-temperature structural applications. However, current RCCAs do not possess the high-temperature oxidation resistance required to survive in oxidizing environments for more than a few hours at or above 1000\textdegree C, without relying primarily on an environmental barrier coating. Here, we present a machine-learning framework designed to predict the oxidation-induced specific mass changes of RCCAs exposed for 24 h at 1000$^\circ$C in air, in order to support the search for oxidation-resistant alloys over a wide range of compositions. A database was constructed of experimental specific mass change data, upon oxidation at 900-1000$^\circ$C for 24 h in air, for 77 compositions comprised of simple elements, binary alloys, and higher-order elemental systems. We then developed a Gaussian Process Regression (GPR) model with physics-informed descriptors based on oxidation products, capturing the fundamental chemistry of oxide formation and stability. Application of this GPR model to the database yielded a MAE (mean absolute error) test score of 5.78 mg/cm\(^2\), which was a significant improvement in accuracy relative to models only utilizing traditional alloy-based descriptors. Our model was used to screen over 5,100 quaternary RCCAs, revealing compositions with significantly lower predicted specific mass changes compared to existing literature sources. Overall, this work establishes a versatile and efficient strategy to accelerate the discovery of next-generation RCCAs with enhanced resistance to extreme environments.
\end{abstract}
\newpage
\section{Introduction}

Refractory Complex Concentrated Alloys (RCCAs), a subset of high entropy alloys (HEAs), are considered potential candidates for replacing Ni-based superalloys and conventional refractory alloys due to their enhanced mechanical properties at high temperatures \cite{Muller2019OnAlloys,Miracle2020RefractoryRSAs, Senkov2010RefractoryAlloys, Xiong2023RefractoryProperties}. RCCAs typically contain at least four elements, with each element possessing a concentration varying from 5-35 at\% (unlike traditional alloys).  These alloys tend to occupy the central concentration ranges of phase diagrams and, with multiple principal elements, can span a vast compositional space \cite{Murty20191Alloys, Ye2016High-entropyProspects, Cantor2021MulticomponentAlloys, Yeh2013AlloyAlloys}.  Even relatively simple assessments of room-temperature mechanical properties of RCCAs over such a large compositional space remain intractable using traditional methods, and evaluation of complex characteristics at high temperatures, such as resistance to fatigue, creep, and oxidation, is even less feasible. 

A key focus of prior RCCA research has been the identification of atomic and microstructural factors that could influence the mechanical properties of various refractory phases. Several studies have characterized such refractory phases using parameters such as atomic size mismatch ($\delta$), enthalpy of mixing ($\Delta$$H_{mix}$), entropy of mixing ($\Delta $$S_{mix}$), and various dimensionless quantities\ \cite{Bejjipurapu2021DesigningLearning, Qu2019TheAlloys, Dai2020UsingAlloys, Islam2018MachineAlloys, Li2019Machine-learningAlloys, Pei2020Machine-learningRules, Huang2019Machine-learningAlloys}. While researchers have begun to use such factors to assess RCCA compositions optimized for mechanical and other thermo-physical properties \cite{Hastings2025AcceleratedSpace, Khatamsaz2023BayesianApproach, Arroyave2022ADesign, Gou2024Multi-objectiveLearning}, there has been notably less exploration into developing guidelines for RCCAs with enhanced resistance to high-temperature oxidation. Initial experimental investigations of the oxidation of RCCAs have often been conducted by introducing new elements into base refractory metals or alloys, adjusting the compositions of such alloys, and then developing correlations between alloy composition and oxidation behavior 
\cite{CAO2019EffectsAlloys, Gorr2020ATemperatures, Schellert2021TheTa-Mo-Cr-Ti-xAl}. While these strategies have led to dramatic improvements in oxidation resistance for certain alloy systems \cite{Gorr2017High-TemperatureComposition, GorrBronislavaandMueller2018DevelopmentStrategy, Lo2022ElementalAlloys, Schellert2021TheTa-Mo-Cr-Ti-xAl}, these RCCAs still possess specific mass gain rates that are notably higher than for state-of-the-art Ni-based superalloys. Furthermore, although this experimental approach can be a satisfactory means of identifying compositions within a specific RCCA system with enhanced oxidation resistance, it is not amenable for evaluation of the oxidation behavior of a wide range of RCCA compositions \cite{Giles2022Machine-learning-basedStrength}.

Prior efforts to predict the oxidation behavior of RCCAs and HEAs have primarily employed composition-based machine learning (ML) approaches. Bhattacharya \textit{et al.} \cite{Bhattacharya2020PredictingLearning} developed regression models using alloy compositions to estimate the parabolic rate constants of Ti-based alloys. Taylor \textit{et al.} \cite{Taylor2021HighLearning} incorporated activation energy as a descriptor for oxidation resistance and compared several ML algorithms, including linear regression, random forest, single decision tree, artificial neural network, and k-nearest neighbor, using datasets spanning steels, superalloys, and aluminides.  

Several studies have further explored neural-network and hybrid ML frameworks. Cui \textit{et al.} \cite{Cui2021MachineSteels} trained a back-propagated neural network and a support vector machine to predict oxide scale thickness and deformation in steels. Dong \textit{et al.} predicted the oxidation resistance of Al–Cr-containing alloys using a random forest model, while Dewangan \textit{et al.} \cite{Dewangan2022ApplicationAlloys} utilized an artificial neural network (ANN) model incorporating alloy composition, exposure time, and oxidation temperature to predict high-temperature oxidation-induced mass gains in Al–Cr–Fe–Mn–Ni–W HEAs. The ANN model exhibited excellent predictive accuracy, with a Pearson correlation coefficient exceeding 0.999 between predicted and experimentally measured mass gains.  

In parallel, automated and high-throughput ML strategies have been employed to enhance predictive capability. Loli \textit{et al.} \cite{Loli2022PredictingMethods} implemented an automated ML pipeline (“Tree-based Pipeline Optimization Tool”) to predict mass change values at 1000\textdegree C across various multi-principal element alloys. Kim \textit{et al.} \cite{KimRegressionNetwork} and Duan \textit{et al.} \cite{Duan2023DesignLearning} applied ANN models to Ni-based superalloys (Ni–Co–Cr–Mo–W–Al–Ti–Ta–C–B and Ni–Co–Cr–Mo–Al–Ti–Nb–Hf–Zr systems, respectively) using proprietary or experimentally derived datasets. However, these models were limited to interpolation within known composition spaces, as correlations for unseen alloy compositions were not readily generalizable.  

Overall, while these studies demonstrate the potential of ML approaches for oxidation prediction, they largely depend on composition-only descriptors. This limitation motivates the present work, which incorporates thermodynamic and oxide-related descriptors to capture the underlying physical mechanisms governing oxidation resistance.
    
Historically, the development of alloys that resist oxidation has been based on the capability of such alloys for forming protective oxide scales on the alloy surface; that is, the formation of a dense, inert, adherent, slowly-thickening external oxide scale has been crucial for achieving a low rate of alloy oxidation at a high temperature.  The composition, structure, and integrity of the oxide product(s) formed during the oxidation of an RCCA can be influenced by several factors, including the chemical composition of the alloy, the exposure atmosphere, the temperature and duration of such exposure, and the stresses resulting from oxide scale thickening (growth stresses) and/or from thermal expansion mismatches between the oxide product(s) and the underlying alloy (thermal stresses) \cite{Xu2000Pilling-bedworthAlloys}. The oxidation resistance of an RCCA can be strongly influenced by characteristics of the oxide product(s) formed during oxidation, as well as by inherent properties of the alloys.  However, prior work to predict the oxidation behavior of RCCAs has not been highly focused on the development of key descriptors related to specific characteristics of the oxidation products.  The objective of this paper is to identify new oxidation product-based descriptors that improve predictive models of specific mass gain, in order to enable the rapid discovery of new oxidation-resistant RCCAs. Our predictive model has been applied in this work to a relatively large design space comprising 84 unique quaternary aluminum-containing alloy systems.

\section{Methods}
\begin{figure}[H]
  \centering
  \includegraphics[width=\textwidth]{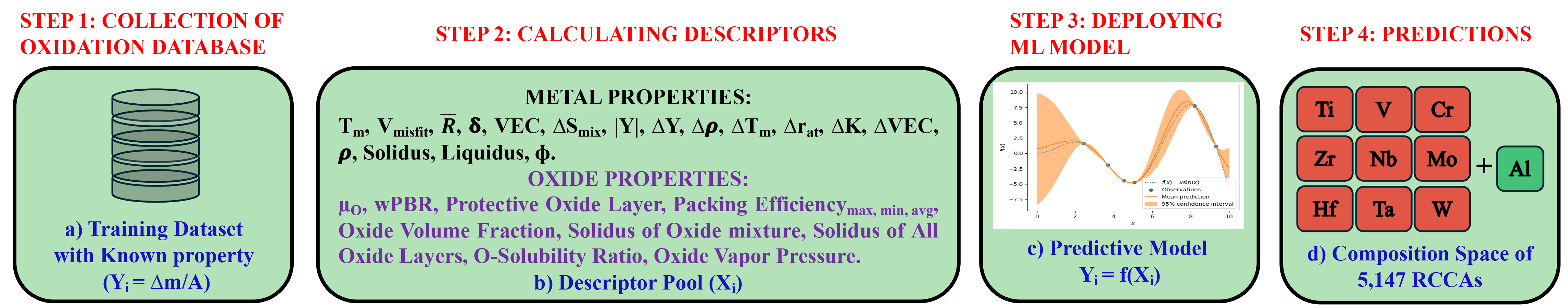}
  \caption{
   Integrated machine learning framework for predicting the high-temperature oxidation behavior of Refractory Complex Concentrated Alloys (RCCAs)}
  \label{fig:ML}
\end{figure}
Figure~\ref{fig:ML} illustrates the overall process for developing the ML (machine learning) oxidation model, including data collection and curation, calculation of descriptors associated with key properties of the metal alloy and of the oxidation product(s), training of ML models, and prediction and experimental validation of the oxidation resistance of new compositions. 
\subsection{Data Collection and Curation}
The initial oxidation dataset for this study, obtained from Mishra \textit{et al.} \cite{Mishra2024MassSharing}, contained specific mass gain data for the oxidation of pure elements, binary alloys, ternary alloys, and MPEAs (defined here as alloys with four or more principal elements). The entire dataset comprised 408 data points from 145 unique compositions collected from 68 literature sources at different temperatures. The oxidation data were analyzed and indexed using the FAIR (Findable, Accessible, Interoperable, and Reusable) database tool \cite{Saswat2022}, which employs Bayesian calibration to characterize mass uptake versus time behavior and uses Bayesian information criteria to identify the best-fit kinetic model for each oxidation dataset. As shown in SI Figure 1(a), the present study focused on the 77 alloys that underwent oxidation experiments in the temperature range of 900–1000\textdegree C. SI Figure 1(b)  shows the distribution of specific mass gain values (the values of mass change per specimen surface area, $\Delta m/A$) for the 77 alloys tested from 900--1000$^\circ$C. The specific mass gain values spanned 3 orders of magnitude, which indicated the wide variability in oxidation resistance across the different alloy chemistries. These alloy chemistries involved 18 different elements: Al, Ti, V, Cr, Zr, Nb, Mo, Hf, Ta, W, Si, Ni, Co, Fe, Cu, and Mn. The metric employed for ML training and testing was the specific mass gain (in mg/cm\textsuperscript{2}) for each alloy after 24 h at each documented oxidation temperature.

SI Figure 2(a) reveals the distribution of different elements in the 77 alloys oxidized at 900–1000\textdegree C, where the number of counts refers to the number of alloys that contained a given element.  SI Figure 2(b) provides a box plot of the atomic fraction distribution for each element. 

\subsection{Descriptor Calculations}
In the present study, values of descriptors were calculated for each composition in the collected dataset and design space (Section \ref{sec:design_space}) to predict the specific mass change of RCCAs during isothermal oxidation. Two sets of descriptors were utilized: descriptors based on characteristics of the metal alloys (SI Table 1) and descriptors based on characteristics associated with the oxide products (Table 1).

For the descriptors associated with metal alloy characteristics, $c_{i}$ refers to the atomic percentage of the $i$th element in the RCCA. The first 7 of these alloy-based descriptors were calculated using the rule of mixtures. The next 6 descriptors were calculated based on the difference between the maximum and minimum property values of the corresponding elements in the alloy. The last 4 of these alloy-based descriptors were the CALPHAD-calculated properties of bulk density, solidus temperature, liquidus temperature, and an encoding of the phase information at 1000\textdegree C of the alloy (as one-hot encoding that included BCC phases, FCC phases, BCC + secondary phases, and FCC + secondary phases). All CALPHAD-based calculations were conducted using Thermo-Calc® 2023a software, with the TCHEA6.0 database implemented via the TC-Python API \cite{The2023a,Bejjipurapu2025PyTCPlotter, Karumuri2023HierarchicalMeasurements}.

\begin{figure}[H]
  \centering
  \includegraphics[width=\textwidth]{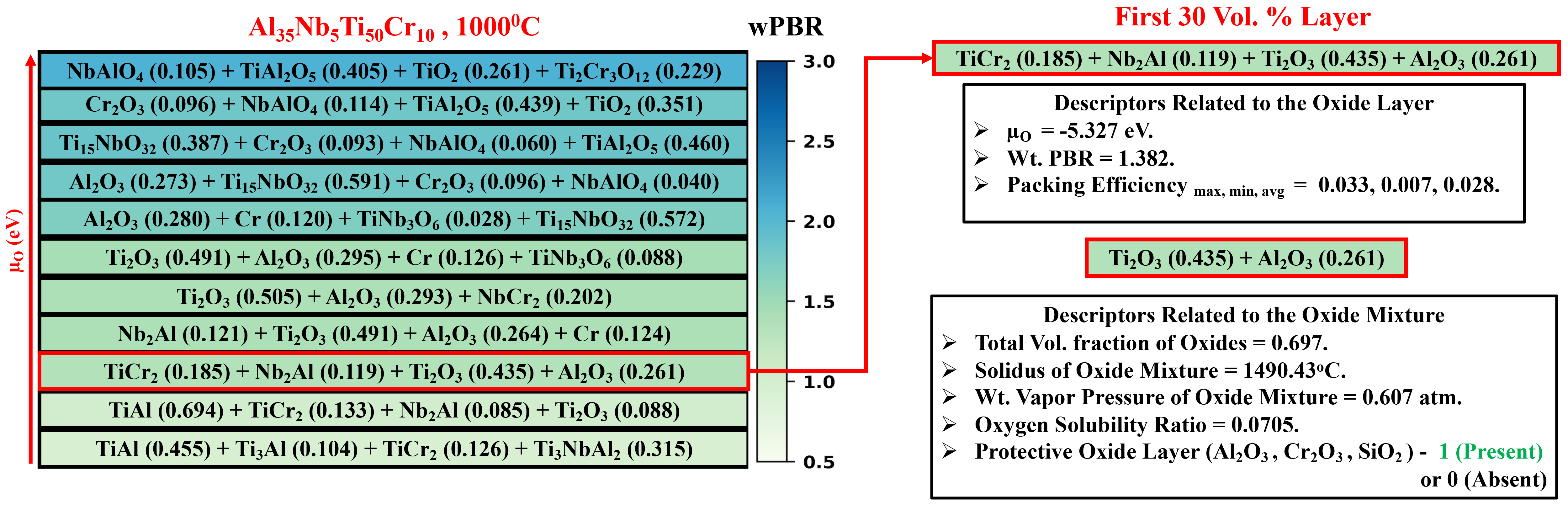}
  \caption{Left: Prediction of the multi-layered structure of the oxidation products formed upon oxidation of a Al\textsubscript{35}Nb\textsubscript{5}Ti\textsubscript{50}Cr\textsubscript{10} alloy at 1000$^{\circ}$C in air, with the most oxygen-rich layer (highest oxygen chemical potential) shown at the top. Right: For the first layer containing 30 vol percent oxide, descriptors are shown for characteristics associated with the entire layer and for characteristics associated with the oxide mixture in this layer. } 
  \label{fig:Layers}
\end{figure}

In order to provide descriptors associated with the oxides formed upon alloy oxidation, a thermodynamics-based model was first used to predict the products of such oxidation. The oxidation products were assumed to form in a multi-layered structure, with the phase content in each layer determined from a model developed by Butler \textit{et al.} \cite{Butler2022OxidationAlloys, L.G.WarePyOxidation} based on a database of \textit{ab initio} calculations. It was assumed in this model that the system was closed with respect to metal atoms (i.e., no metal loss, such as by volatilization) and open with respect to oxygen (i.e., an oxygen chemical potential gradient was allowed to develop across the multiple layers). The phase content within a given layer was determined for the range of values of oxygen chemical potential, \(\mu_O\), within that layer by minimizing the oxygen grand potential, \(\overline{\Phi}_G\), given by:

\begin{equation}
\overline{\Phi}_G (\mu_O, N_M, P, T) = \frac{E_0 - \mu_O N_O}{N_M}
\end{equation}

\noindent where \( E_0 \) is the DFT-calculated formation energy at 0 K, \( \mu_O \) is the chemical potential of oxygen, \( N_M \) is the number of metal atoms per formula unit, \( N_O \) is the number of oxygen atoms per formula unit, \( P \) is the pressure, and \( T \) is the temperature.  The formation energy for each phase was retrieved from the Materials Project database \cite{Jain2013Commentary:Innovation}. An example of the model output for the multiple layers formed upon oxidation of a \ce{Al35Nb5Ti50Cr10} alloy at 1000\textdegree C in air is shown in Figure \ref{fig:Layers} (note: the gradation in color in this figure indicates the change in values of the weighted Pilling-Bedworth Ratio, wPBR, described below).

With this model, the number of layers predicted to form upon oxidation of a given alloy depended on the various oxidation products that could form at different oxygen chemical potentials. The varying number of such layers for different alloys complicated calculation of standardized oxidation product-based descriptors. Arroyave \textit{et al.} \cite{Sauceda2022HighAlloys} utilized a singular metric, the Area-Under-the-Curve-2 (AUC2), to characterize oxidation resistance by integrating phase fraction and chemical potential as a function of temperature and oxygen partial pressure. We instead focused on the development of descriptors associated with the first layer that contained a total oxide volume fraction greater than 0.30, which was denoted as the \FTH{} This choice was associated with Wagner’s theory \cite{Wagner1959ReaktionstypenLegierungen} for the transition from internal to external oxidation of an alloy. Wagner proposed that this transition should occur when a critical volume fraction of internal oxide particles forms at the oxidation front. At and beyond this critical oxide volume fraction, denoted as \( f_v^* \), the sideways growth of existing internal oxide particles at the oxidation front should be sufficiently favored, over the nucleation of new oxide particles in advance of the existing particles, that such sideways growth should result in the formation of an interconnected oxide layer.  While Wagner did not specify how to calculate \( f_v^* \), later experimental work by Rapp yielded a \( f_v^* \)  value of approximately 0.3 for Ag–In alloys undergoing oxidation at 550$^\circ$C \cite{Rapp1961TheAlloys}. Similar or higher values of\( f_v^* \) have been discussed for other alloy systems undergoing oxidation at higher temperatures\cite{Nesbitt1989PredictingOxidation, Zhao2015QuantitativeOxidation}. By developing descriptors associated with the \FTH{}, our model could evaluate potential alloys for which external oxidation may be possible.
As indicated by the example illustrated In Figure~\ref{fig:Layers}, each layer in the multilayer structure is comprised of a distinct set of phases formed upon oxidation of the alloy at 1000\textdegree C. The \FTH{} is highlighted, and key descriptors associated with this layer (\(\mu_O\), wPBR, and packing efficiency) and with the oxides present in this layer (total volume fraction of oxides, solidus temperature, vapor pressure, oxygen solubility ratio, and the presence or absence of an oxide capable of forming a slow-growing, protective layer) are indicated. Descriptions of these oxidation-based descriptors, and procedures for calculating such descriptors, are described below:
\begin{enumerate}

    \item[1.] \textbf{ Oxygen Chemical Potential }(\(\mu_O\)):  The minimum \(\mu_O\) required to form the phases in \FTH{} was calculated.  This \( \mu_O \) is related to the equilibrium partial pressure of oxygen, \(pO_{2}\) , at 1000$^{\circ}C$ required for the formation of the oxides in the \FTH{}, as described by Reuter \textit{et al.} \cite{Reuter2002CompositionPressure}. 

    \item[2.] \textbf{Weighted Pilling-Bedworth Ratio }(wPBR)\textbf{:} The Pilling-Bedworth Ratio (PBR) is the ratio of the volume of the elementary cell of a metal oxide to the volume of the elementary cell of the metal consumed to form the metal oxide, and has been used to evaluate whether a dense, compact, external metal oxide could form on a metal surface as well as for consideration of stresses that may develop in the metal oxide scale\cite{Xu2000Pilling-bedworthAlloys, Liu2023DesigningMechanisms, pilling1923oxidation} . Since multiple oxide and metal phases were present in the \FTH{} for the RCCAs evaluated in the present work, a weighted PBR (wPBR) descriptor was developed. The wPBR value was defined as the ratio of the total volume of the oxidation products (all oxide and metal phases) in the \FTH{} , weighted by their volume fractions and atomic volumes, to the total volume of the underlying layer.  The wPBR was expressed mathematically as  \cite{Butler2022OxidationAlloys}:
    \begin{equation}
        \text{wPBR} = \frac{\sum \left(X_i \frac{V_i}{N_m} \right)}{V_s}
    \end{equation}

    where \( X_i \) is the volume fraction of the $i$th phase, \( V_i \) is the volume per unit cell of the phase, \( N_m \) is the number of metal atoms in the unit cell, and \( V_s \) is the weighted average volume per atom of the substrate. 

    \item[3.] \textbf{Protective Oxide in the Layer }(binary descriptor)\textbf{:} \ce{Al2O3}, \ce{Cr2O3}, and \ce{SiO2} are capable of forming as slow-growing, protective, external scales under appropriate oxidation conditions on metal alloys containing sufficient amounts of Al, Cr, and/or Si, respectively \cite{Brady2000AlloyFormation}. A binary descriptor was used to indicate the presence (1) or absence (0) of at least one of these oxides within the \FTH{} 

    \item[4-6.] \textbf{Average, Maximum, and Minimum Packing Efficiency:} The rate of ionic diffusion through an oxide can be affected by the relative packing density of the oxide structure  \cite{Brumm1992TheAlloys}.  The average packing efficiency of the oxide(s) in the \FTH{} by \cite{Butler2022OxidationAlloys} was calculated as follows:
    
    \begin{equation}
        \text{Packing Efficiency}_{\text{avg}} = \sum (X_i \cdot \text{PE}_i)
    \end{equation}
    
    where \( X_i \) is the volume fraction of oxide phase i, and \( \text{PE}_i \) is the packing efficiency of oxide phase i. In addition to the average value of the packing efficiency, the maximum and minimum values of the packing efficiency were also calculated for individual oxides present in the \FTH{} 

    \item[7.] \textbf{Volume Fraction of Oxides:} The total volume fraction of the oxide phases in the \FTH{}  was calculated to quantify the relative extent of oxidation in this specific layer for each RCCA.

    \item[8, 9.] \textbf{Solidus Temperature of the Oxide Mixture in One and All Layers }(binary descriptor)\textbf{:} The solidus temperature of the oxide mixture in the \FTH{} was used to identify oxide mixtures that could form liquid at or below 1000$^\circ$C, which could degrade oxidation resistance. The solidus temperature of the oxide mixture in this layer was calculated using the Thermo-Calc\textsuperscript{\textregistered} 2023a software with the TCOX11.0 database, accessed via the TC-Python API for non-Hf- and -Ta-containing alloys \cite{The2023a} as such alloys were not available in this database. For Hf and Ta-containing alloys, a high-temperature property model developed by Zachary et al. \cite{McClure2021ExpandingSelection} was used instead. 
    A binary descriptor (0, 1) was also used to determine if the solidus temperature of the oxide mixture in all layers was below 1000$^\circ$C, with zero (0) indicating that no layer contained an oxide mixture with a solidus temperature below 1000\textdegree C, and one (1) indicating that at least one layer possessed an oxide mixture with a solidus temperature below 1000\textdegree C.
    
    \item[10.] \textbf{Oxygen Solubility Ratio} ($\frac{N_O}{N_B}$)  Alloys with high values of oxygen solubility can be susceptible to internal oxidation \cite{Guan1994OxygenAlloys}. Within the compositional space examined in this study, the constituent elements span a wide range of oxygen solubilities at 1000\textdegree C. The group IV elements, Ti, Zr, and Hf, exhibit particularly high oxygen solubilities at 1000\textdegree C, with values in excess of 30 at\% for $\alpha$-Ti \cite{fischer1997thermodynamic}, 25 at\% for $\alpha$-Zr \cite{abriata1986zr}, and 15 at\% for $\alpha$-Hf \cite{domagala1965hafnium}, which makes alloys rich in these elements particularly vulnerable to internal oxidation. The group V elements, V, Nb, and Ta, exhibit significant but lower oxygen solubilities of about 8 at\% \cite{stringer1965vanadium}, 2.5 at\% \cite{jehn1972high}, and 3 at\% \cite{massih2006thermodynamic}, respectively, at 1000\textdegree C. The group VI elements, Cr, Mo, and W, exhibit much lower oxygen solubilities (below 0.05 at\%) \cite{Gorr2021CurrentAlloys}. In this study, the oxygen solubility ratio ($\frac{N_O}{N_B}$) was used as a descriptor, with $N_O$ referring to the mole fraction of oxygen dissolved in the alloy and $N_B$ referring to the total mole fraction (sum) of the non-oxygen elements in the alloy. The oxygen solubility values for non-Hf and -Ta-based alloys were calculated using Thermo-Calc\textsuperscript{\textregistered} 2023a software with the TCOX11.0 database, accessed via the TC-Python API \cite{The2023a}. For Hf- and/or Ta-containing alloys, a rule-of-mixtures approach was used to estimate oxygen solubility, accounting for the contribution of each element based on mole fraction of that element in the alloy. 

    \item[11.] \textbf{Vapor Pressure of Oxide Mixture:} A high rate of vaporization of an oxide scale can, at the very least, reduce the scale thickness and degrade the protective nature of the scale. Certain oxides, notably those of the group VI metals (such as \ce{MoO3}, \ce{WO3}, and \ce{Cr2O3}), exhibit relatively high vapor pressures and vaporization rates in air at 1000\textdegree C. \cite{Gorr2021CurrentAlloys}.  A rule-of-mixtures approach was utilized, based on the mole fraction and vapor pressure of each oxide in the \FTH{} at 1000\textdegree C, to calculate the \textit{Weighted Vapor Pressure} descriptor, as follows:
        \[
        P_{\text{weighted}} = \sum_{i} x_i \cdot P_i
        \]

    where $x_i$ is the normalized mole fraction of oxide $i$ (estimated from the thermodynamic equilibrium oxide phase fractions), and $P_i$ is the vapor pressure of oxide i at 1000\,$^\circ$C. Values of oxide vapor pressure at 1000\textdegree C were compiled from open-source literature \cite{article}. 

\end{enumerate}
\renewcommand{\arraystretch}{2.0}
\begin{longtable}{@{}c>{\raggedright\arraybackslash}p{2.0cm}>{\centering\arraybackslash}m{5.0cm}>{\raggedright\arraybackslash}p{6.0cm}@{}}
\caption{Summary of all 11 oxidation product-based descriptors developed in this study.}
    \label{tab:oxidation_descriptors} \\
    
    \toprule
    \textbf{Index} & \textbf{Notation} & \textbf{Formulation} & \textbf{Descriptor Description} \\
    \midrule
    \endfirsthead
    
    \toprule
    \textbf{Index} & \textbf{Notation} & \textbf{Formulation} & \textbf{Descriptor Description} \\
    \midrule
    \endhead
    
    \bottomrule
    \endfoot
    
    1 & $\mu_O$ & -- & Minimum chemical potential of oxygen at which there is a mixture of oxidation products with a total oxide volume fraction $>$ 0.30 (\FTH{}). \\
    2 & wPBR & -- & Weighted Pilling--Bedworth Ratio of the \FTH{}. \\
    3 & Protective Oxide Layer & 0 / 1 & Absence (0) or presence (1) of \ce{Al2O3}, \ce{Cr2O3}, or \ce{SiO2} in the \FTH{}. \\
    4--6 & Packing Efficiency$_{\text{max,min,avg}}$ & -- & Max, min, and average packing efficiency among oxides in the \FTH{}. \\
    7 & Oxide Volume Fraction & -- & Sum of volume fractions of oxides in the \FTH{}. \\
    8 & Solidus of Oxide Mixture & -- & Solidus temperature of the oxide mixture in the \FTH{}. \\
    9 & Solidus of All Oxide Layers & 0 / 1 & Absence (0) or presence (1) of any oxide layer with a solidus temperature $<$ 1000\textdegree C. \\
    10 & Oxygen Solubility Ratio & 
    $\frac{N_O}{N_B} = \frac{\text{Oxygen solubility}}{\sum\limits_{i \neq O} \text{composition}_i}$ & 
    Ratio of oxygen solubility (mole fraction) in the alloy (N$_O$) to the sum of the mole fractions of the oxidizing elements in the bulk alloy (N$_B$). \\
    11 & Oxide Vapor Pressure & -- & Weighted vapor pressure of the oxide mixture in the \FTH{} at 1000\textdegree C. \\
    
\end{longtable}

\subsection{Machine Learning for Specific Mass Change}

A Gaussian Process Regression was used with a radial basis function (RBF) kernel and Automatic Relevance Determination (ARD), which allowed for a separate length scale for each descriptor dimension. This approach used the model to automatically determine the relevance of each descriptor, which further enhanced the adaptability of the GPR to complex datasets. By automatically determining the relevance of each descriptor, the ARD-RBF kernel could effectively handle high-dimensional input spaces, which made the GPR a suitable choice for a wide range of descriptors. 

The initial dataset was split into training and testing sets, with a training-to-testing ratio of 0.80 to 0.20. The performance of the GPR model was assessed using the value of the Mean Absolute Error (MAE). The MAE value of the machine learning model (GPR) was determined using repeated stratified \( k \)-fold cross validation (\( k = 5 \)). This approach helped to mitigate overfitting issues that could arise from smaller datasets. With this process, the entire dataset was randomly divided into five smaller sets (folds). The model was trained using four folds for each of the five iterations and then validated using the remaining single fold. This procedure was repeated three times with different random splits, which resulted in the generation of fifteen sets to be used to estimate the performance of the model.

\subsection{Alloy Design Space}
\label{sec:design_space}

\begin{figure}[H]
  \centering
  \includegraphics[width=\textwidth]{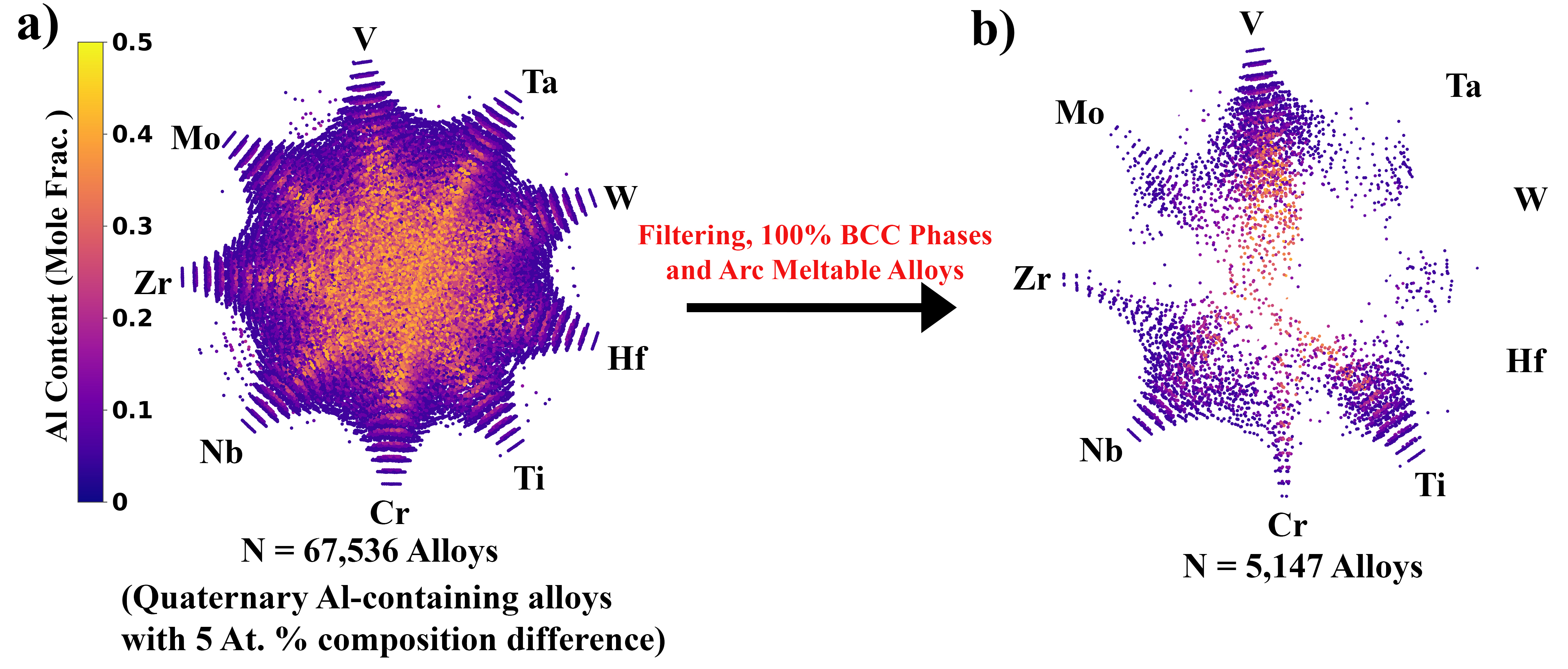}
  \caption{(a) Full design space (mapped into 2D via MDS) and (b) design space restricted to alloys with only BCC phases processable via arc melting.}
  \label{fig:MDS}
\end{figure}

The initial design space consisted of quaternary alloys comprised of Al combined with three refractory elements chosen from Ti, Zr, Hf, V, Nb, Ta, Cr, Mo, and W. Composition step increments of 5 atomic percent (at\%) were used, with the Al content limited to a range of 5–40 at\% in order to reduce intermetallic phase formation. Our process generated 67,536 unique alloys spanning 84 different quaternary alloy systems, as illustrated by the Multi-Dimensional Scaling (MDS) plot in Figure~\ref{fig:MDS}(a) . This design space was narrowed to alloys suitable for fabrication by arc melting, by selecting compositions with predicted liquidus temperatures (without Al and Cr additions) below the boiling points of Al and Cr (which were 2518\textdegree C and 2669\textdegree C, respectively \cite{Barin1995ReferenceBar}). A second filter for only body-centered-cubic (BCC) phases was then applied. After applying the arc melting and BCC phase filtering criteria, the composition space was reduced to 5,147 alloys, as shown in the second MDS projection in Figure~\ref{fig:MDS}(b).

\subsection{Distributions of Oxide-Related Descriptors Across Alloy Design Space}
\begin{figure}[H]
    \centering
    \includegraphics[width=\textwidth]{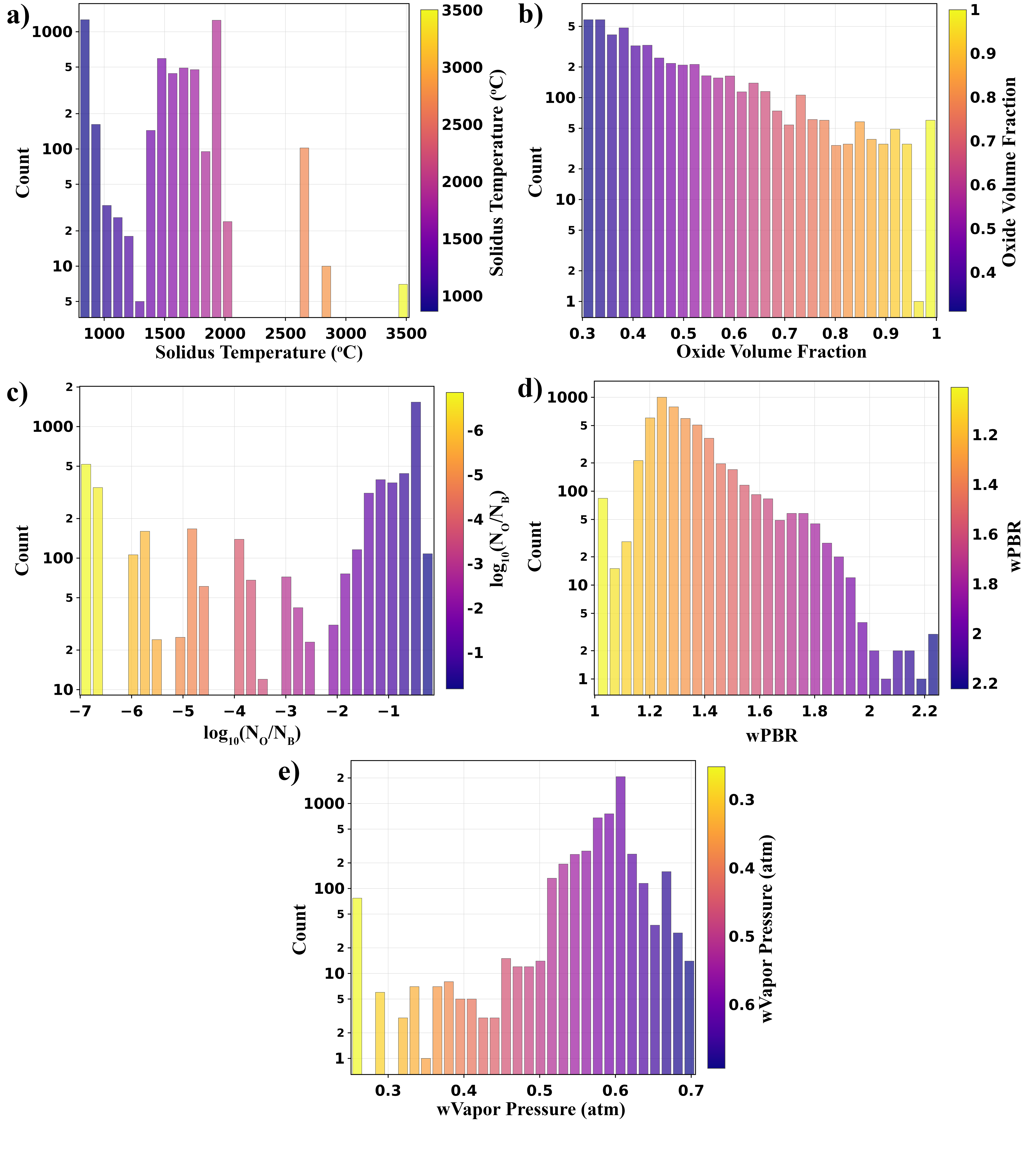}
    \caption{
    Distributions of key oxide-related descriptors used in the oxidation resistance framework:  
    (a) Oxide solidus temperature (\textdegree C), (b) Oxide volume fraction, 
    (c) log$_{10}$($\frac{N_O}{N_B}$), (d) Weight Pilling-Bedworth Ratio (wPBR), 
    and (e) Weighted oxide vapor pressure (atm) visualized across alloy design space.}
    \label{fig:property_histograms}  
\end{figure}

The histograms shown in Figure~\ref{fig:property_histograms} provide the distributions, across the alloy design space, of the key oxidation-related descriptors calculated in the \FTH{}. Each descriptor was color mapped, with yellow representing the most favorable descriptor values for resistance to oxidation. The histograms quantified, across the design space,  how frequently favorable values occurred for each descriptor. For a majority of the alloys, the \FTH{} contained high-melting oxide mixtures (for \(72.0\%\) of alloys, \(T_\mathrm{solidus}>1000^\circ\mathrm{C}\)) and possessed values of wPBR that fell in the acceptable range of 1-2 (for \(99.8\%\) of alloys), as shown in Figures~\ref{fig:MDS}(a) and (d), respectively. However, only for a minority of the alloys (\(36.1\%\)) did the \FTH{} contain an oxide volume fraction \(>0.5\) (\
(Figure~\ref{fig:MDS}(b)), which is typically needed for the formation of continuous external scales.  Because the oxygen solubility ratio spanned several orders of magnitude (Figure~\ref{fig:MDS}(c)), the logarithm of this ratio, \(\log_{10}(N_\mathrm{O}/N_\mathrm{B})\), was used in both the histogram and heat map to aid in visualization and interpretation. For a majority of the alloys (\(65.7\%\)), the \(\log_{10}(N_\mathrm{O}/N_\mathrm{B})\) values fell in the undesired 0 to -2 range (i.e., (\(N_\mathrm{O}/N_\mathrm{B})\) values exceeded 0.01), as shown in Figure~\ref{fig:MDS}(c). For the vast majority of alloys (\(96.7\%\)), the  \FTH{} contained oxide mixtures with undesired, high oxide vapor pressures exceeding 0.5 atm (Figure~\ref{fig:MDS}(e)). Taken together, these histograms indicated that all of the desirable descriptor values rarely occurred for the same alloy. Most alloy compositions exhibited some, but not all, desirable values which indicated the need to consider trade-offs between the various oxidation product-based descriptors.

\begin{figure}[H]
    \centering
    \includegraphics[width=\textwidth]{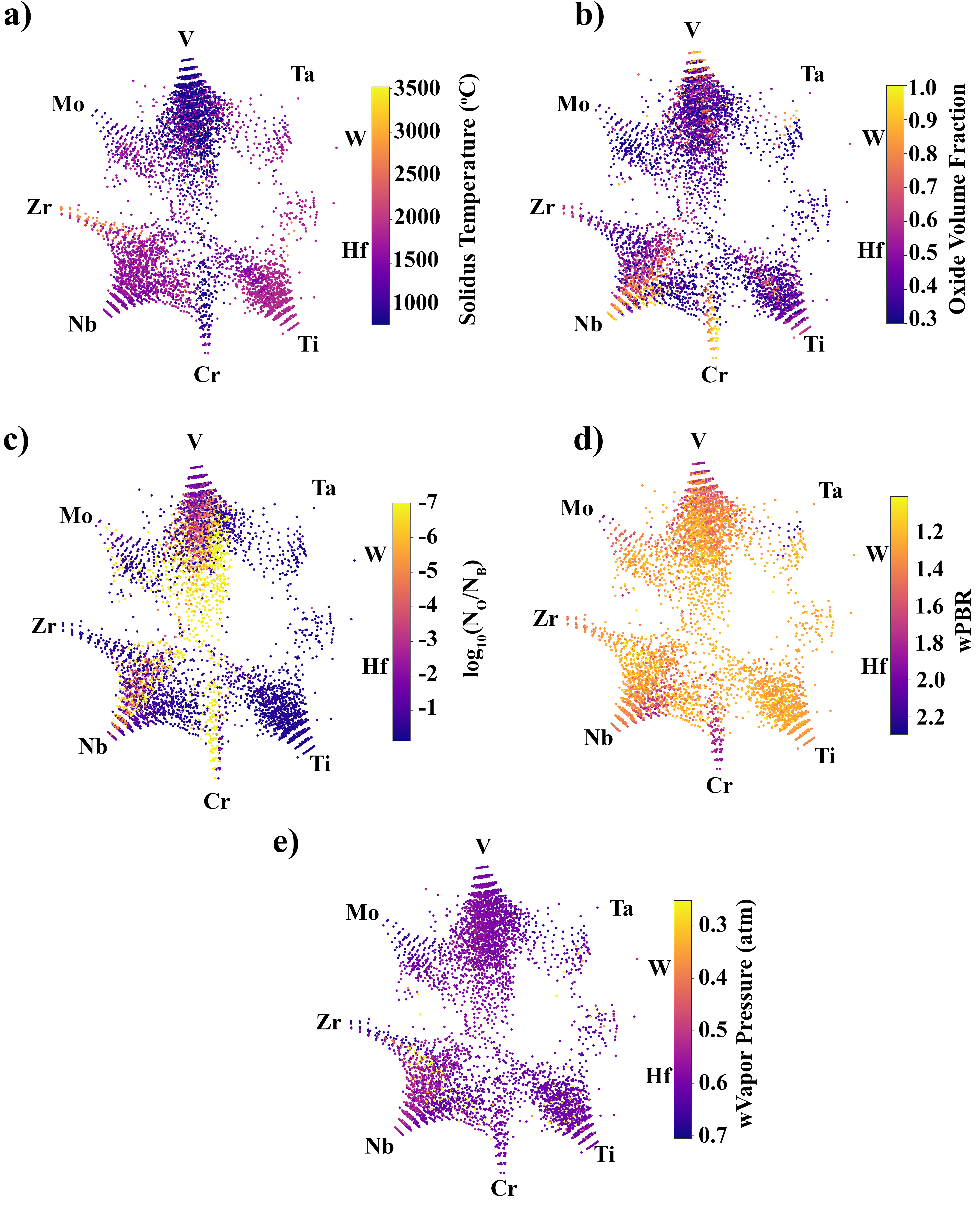}
    \caption{Property distributions in high-dimensional alloy space where a strong yellow color represented favorable values for improved oxidation resistance. 
    (a) Oxide solidus temperature (\textdegree C), (b) Oxide volume fraction, 
    (c) log$_{10}$($\frac{N_O}{N_B}$), (d) Weight Pilling-Bedworth Ratio (wPBR), 
    and (e) Weighted oxide vapor pressure (atm) visualized across alloy design space.}
    \label{fig:property_distributions}
\end{figure}
The heatmaps in Figure~\ref{fig:property_distributions} reveal the influences of alloy composition on the values of the oxidation product-based descriptors, with each descriptor color mapped using the same color scale as shown in the histograms in Figure 4. Desired values of these descriptors for optimal oxidation resistance included a high solidus temperature for the oxide mixture in the \FTH{} (Figure~\ref{fig:property_distributions}(a)), a high oxide volume fraction in this layer (Figure~\ref{fig:property_distributions}(b)) for continuous oxide scale formation, a low oxygen solubility ratio (Figure ~\ref{fig:property_distributions}(c)) to minimize internal oxidation, a Wt.\ PBR value in the range of 1--2 (Figure~\ref{fig:property_distributions}(d)) to balance dense, continuous oxide scale formation against high oxidation-induced stress and associated cracking and spallation, and a low weighted oxide vapor pressure for the oxide(s) in the \FTH{}  (Figure~\ref{fig:property_distributions}(e)) to minimize oxide loss due to evaporation. However, the heat maps in Figure~\ref{fig:property_distributions} do not reveal appreciable overlap of compositional regions clearly aligning with all of these optimal oxidation properties. Such a lack of desired overlap underscores the complexity of the challenge to identify oxidation-resistant RCCAs and the associated need for integrated, data-driven, multi-objective optimization (e.g., machine-learning models) to balance the competing factors associated with superior oxidation resistance. By integrating the outputs of the \textit{ab initio}-based model with these descriptors, the present work aims to provide a comprehensive prediction model for oxidation resistance, enabling the design of new alloys optimized for high-temperature applications.

\subsection{SHAP Analysis}
\begin{figure}[H]
    \centering
    \includegraphics[width=\textwidth]{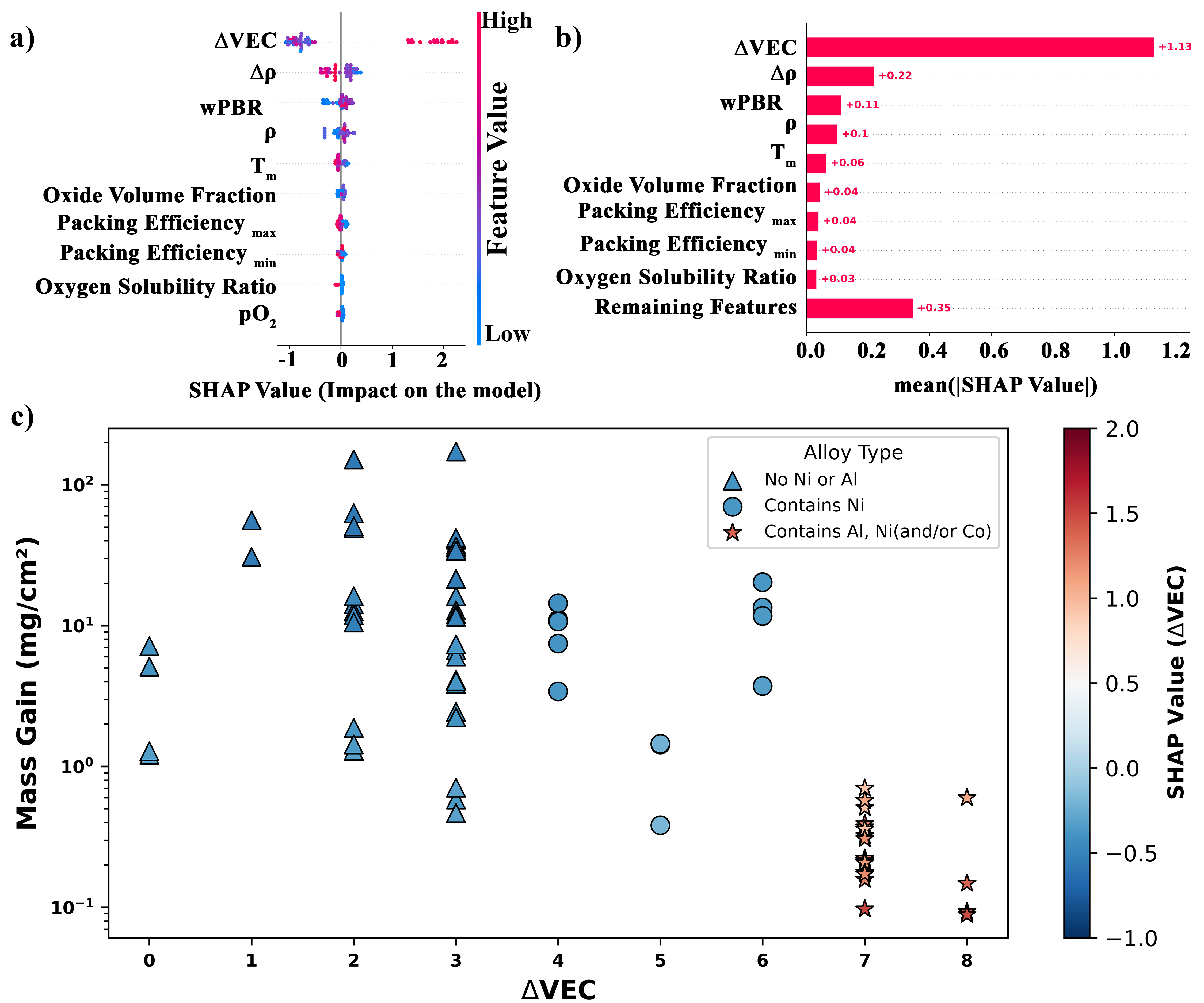}
    \caption{SHAP analysis for determining the importance of descriptors for specific mass gain predictions: (a) SHAP values of the most essential global features for all test set observations. (b) SHAP ranking of the most impactful features in specific mass gain predictions. (c) Relationship between $\Delta$VEC and $\Delta M/A (mg/cm^2)$.}
    \label{fig:SHAP}
\end{figure}

SHAP (SHapley Additive ExPlanations) analysis \cite{Lundberg2017APredictions} was used in this work to evaluate relationships between the calculated input features and oxidation-induced specific mass changes, $-\log(|\Delta M/A|)$. This approach enabled a thorough interpretation of the GPR model by quantifying how each input feature influenced the predictions of the model. Figure~\ref{fig:SHAP}(a) shows the SHAP summary plot, where each descriptor is ranked on the basis of the contribution of that descriptor to the predicted output, $-\log(|\Delta M / A|)$. Points located further to the right corresponded to descriptors that drove the model toward lower values of specific mass gain (higher $-\log(|\Delta M / A|)$). The color gradient in Figure~\ref{fig:SHAP}(a) indicates the magnitude of the descriptor values, with red and blue representing high and low descriptor values, respectively.

Among all 28 descriptors, $\Delta$VEC (the range of valence electron concentration) had the strongest positive impact, followed by $\Delta\rho$, wPBR, and $\rho$. High $\Delta$VEC values (red points) were associated with strongly-positive SHAP values, meaning that larger $\Delta$VEC values resulted in a reduction in the predicted specific mass gain values. This trend was further reflected in Figure~\ref{fig:SHAP}(c), which indicated that alloys with higher $\Delta$VEC values clustered at lower specific mass gain values (the Pearson correlation coefficient for $\Delta$VEC and $-\log(|\Delta M/A|)$ was 0.674). These compositions frequently included Al, Ni, and/or Co, which are known to improve oxidation resistance (note: Ni and Co have relatively high VEC values of 9 and 10, respectively).  \cite{Berthod2005KineticsAlloy, Ju2024UnderstandingCharacterization, Choi1996High-temperatureNi3Al} The contributions of various descriptors towards the mean SHAP values are indicated in Figure~\ref{fig:SHAP}(b). Here, $\Delta$VEC was clearly the dominant predictor of the specific mass gain, with the highest mean absolute SHAP value (+1.13). Other important features, such as $\Delta\rho$, wPBR, and $\rho$, contributed moderately, with mean absolute SHAP values in the range of 0.10-0.22, while the remaining features (N=25) had minimal influence.  ($\Delta$VEC,  $\Delta\rho$, wPBR, $\rho$ + 25 = 29). It is worth noting that five of the top ten most important descriptors identified in the SHAP analysis for predicting the specific mass gain were oxidation product-based descriptors that included 4 descriptors of the  \FTH{} (Wt. PBR, oxide volume fraction, the maximum and minimum oxide packing efficiency values) and the oxygen solubility ratio.

\section{Results and Discussion}

\begin{figure}[H]
  \centering
  \includegraphics[width=\textwidth]{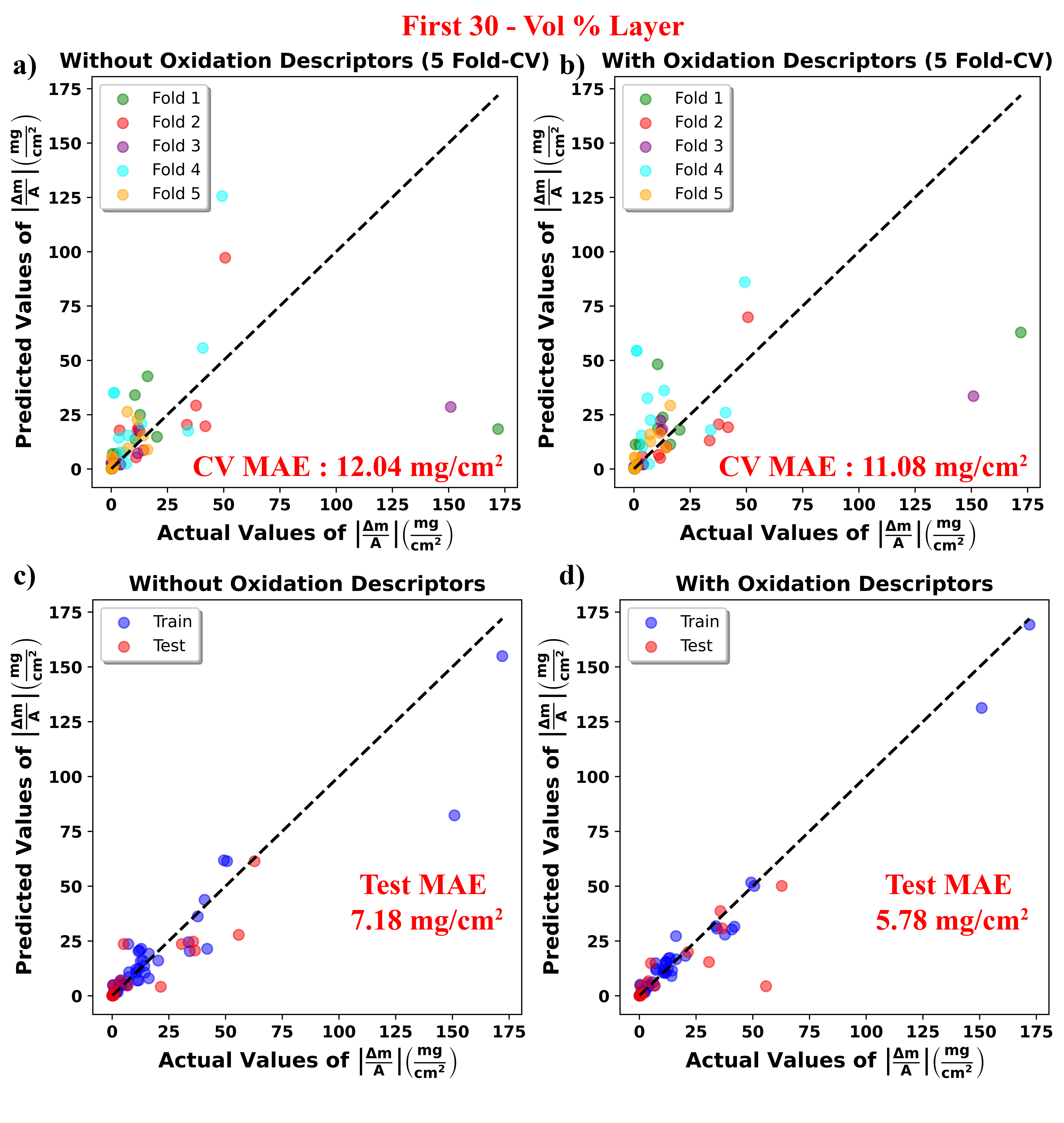}
  \caption{Parity plots comparing predicted vs.\ actual mass change ($|\Delta m|/A$, mg/cm\textsuperscript{2}) for the \FTH{} using the GPR model. 
  (a,b) Results from 5-fold cross-validation without (a) and with (b) oxidation descriptors, showing cross-validation mean absolute error (CV MAE). 
  (c,d) Results for the independent test set without (c) and with (d) oxidation descriptors, highlighting the test MAE improvement when oxidation descriptors are included.}
  \label{fig:GPR}
\end{figure}

The parity plots in Figure~\ref{fig:GPR} reveal the performance of the GPR model, for cases without and with oxidation product-related descriptors, for predicting oxidation-induced specific mass gain \(|\Delta m / A|\) (mg/cm\(^2\)). The baseline model Figure~\ref{fig:GPR}(a) utilized the comprehensive set of alloy-based descriptors detailed in SI Table~1, supplemented by the temperature and partial pressure of oxygen (pO$_{2}$) of the oxidation environment. The enhanced model Figure ~\ref{fig:GPR}(b) incorporated these same alloy-based features along with an additional suite of oxidation product-specific descriptors, as listed in Table ~\ref{tab:oxidation_descriptors}.  Predictions were derived from the transformed target variable \(-\log(|\Delta m / A|)\) and converted back to the linear scale using \(\exp(-\mu + 0.5\sigma^2)\), where \(\mu\) was the predicted mean and with the predicted variance (\(\sigma^2\)) incorporated into the uncertainty estimate (standard deviation) rather than the mean adjustment. Figures~\ref{fig:GPR}(a) and~\ref{fig:GPR}(b) reveal 5-fold cross-validation results, which yielded CV MAE values of 12.04 mg/cm\(^2\)and 11.08 mg/cm\(^2\) for the cases without and with the use of oxidation product-based descriptors, respectively. Figures~\ref{fig:GPR}(c) and~\ref{fig:GPR}(d) display the performance results for the independent test set, where the inclusion of oxidation product-based descriptors again enhanced accuracy, with a reduction in the Test MAE values from 7.18 mg/cm\(^2\) to 5.78 mg/cm\(^2\). These findings indicated that oxidation descriptors reduced scatter around the parity line, thereby improving the ability of the GPR model to capture experimental mass gain trends.

\begin{figure}[H]
    \centering
    \includegraphics[width=\textwidth]{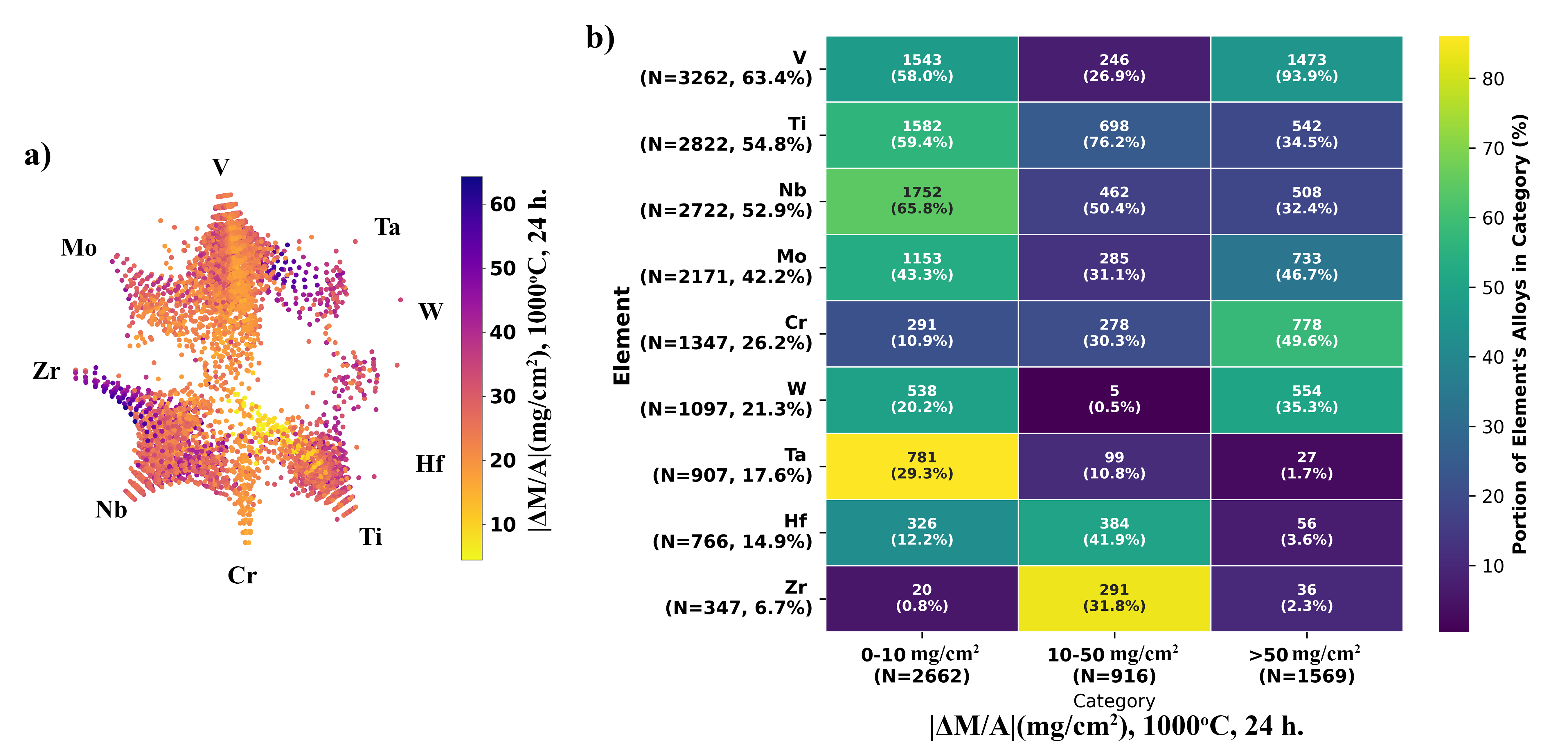}
    \caption{Gaussian Process Regression (GPR) predicted oxidation behavior at 1000\textdegree C for 24 hours. 
    (a) Predicted specific mass gain ($\Delta m/A$) for all alloys in the design space. 
    (b) Labeled heatmap showing the distribution of alloys containing key elements across three mass gain categories. The color of each cell indicates the distribution of a given element's alloys across the categories (normalized by row). The text in each cell provides the absolute alloy count and, in parentheses, the element's prevalence within that category (normalized by column).Labels on the y-axis further detail the total number of alloys (N) containing each element and their percentage of the entire 5,147-alloy design space.}
    \label{fig:predictions}
\end{figure}

The developed GPR model with oxidation product-based descriptors was applied to predict the specific mass changes of alloys in the design space at 1000$^\circ$C over 24 h in air. A 2D-MDS heatmap of the predicted specific mass change values across the compositional design space is provided in Figure~\ref{fig:predictions}(a).  This map reveals that alloys with high concentrations of Zr exhibited relatively high specific mass changes. Zr-rich alloys tend to form \ce{ZrO2} scales, which can be prone to cracking, facilitating further oxidation \cite{Vandegrift2019OxidationOxygen, Lo2019AnOxide}.  This map also reveals that alloys enriched in Cr, Nb, and Ti exhibited lower specific mass gains. Additionally, some alloy compositions located near the center of the MDS plot, representing Al-rich compositions, exhibited relatively low specific mass changes.
To better illustrate compositional trends, alloys in the design space were categorized into three distinct specific mass change groups, as shown in Figure~\ref{fig:predictions}(b) : low specific mass gain (0--10 mg/cm\(^2\)), moderate specific mass gain (10--50 mg/cm\(^2\)), and high specific mass gain ($>$50 mg/cm\(^2\)). Alloys in the low specific mass gain category (N = 2,662) predominantly contained Cr, Ti, V, Mo, and Nb. Alloying with sufficient Cr is well known to provide oxidation resistance, via the formation of an external \ce{Cr2O3} scale,  for Fe-based, Ni-based, and Co-based alloys \cite{Ren2017The850900C, Lu2011OxidationC} Ti and Mo have recently been reported to improve oxidation resistance in certain RCCAs through the formation of mixed ABO\(_4\)-based compounds, particularly when alloyed with Nb, Ta, and Al \cite{Muller2019OnAlloys, Lo2019AnOxide, Schellert2021TheTa-Mo-Cr-Ti-xAl, Gorr2020ATemperatures}, which are also represented in this low specific mass change group. This beneficial alloying effect of Ti with Cr has also been previously reported by Borr et al. \cite{Gorr2020ATemperatures} while studying AlMoCrTaTi alloy systems In contrast, a significant population of V-bearing alloys were found in the high mass change categories. V is well known to form low-melting oxides that are volatile and tend not to be protective \cite{Gorr2021CurrentAlloys}. 

\section{Conclusions}
A set of eleven oxidation product-based descriptors are presented in this study for use in machine-learning frameworks for predicting the oxidation behavior of refractory complex concentrated alloys and high-entropy alloys. These descriptors capture critical aspects of oxidation, including the chemical potential required for oxide formation, the volume change associated with oxidation, ionic permeability through the oxide, oxide melting point, oxide volatility, and oxygen solubility in the alloy. SHAP analysis revealed that the oxidation product-based descriptors were among five of the top ten most impactful for predicting the specific mass gain upon oxidation at 1000$^\circ$C for 24 h in air, although the average valence electron concentration of elements in the alloy was the most influential descriptor for such correlation.
 
The GPR model developed in this work, using these new oxidation product-based descriptors, along with prior thermo-physical-based alloy descriptors, exhibited good predictive capability (MAE of 5.78 mg/cm\(^2\) for testing data) that was enhanced relative to sole use of the thermo-physical-based alloy descriptors (MAE(test) of 7.18 mg/\(^2\)) for estimating the oxidation-induced specific mass change at 1000$^\circ$C for 24 h in air across various RCCA systems. Targeted screening of over 5,147 quaternary RCCAs identified compositions with markedly lower 24 h mass change values at 1000$^\circ$C compared to the training dataset.
\section*{Supplementary Information}
The supplementary material includes distributions of alloy systems, specific mass gain values, and elemental compositions (SI Figure 1 and 2); a complete list of the metal alloy-based descriptors used for the baseline model (SI Table 1); a Pearson correlation heatmap for the oxidation product-based descriptors (SI Figure 3); and supplementary parity plots evaluating the model's performance for an alternative layer definition (SI Figure 4)

\section*{Author contributions}
Akhil Bejjipurapu: Writing – original draft (equal); Data curation (equal); Formal analysis (equal); Investigation (equal); Methodology (equal); Software (equal); Validation (equal); Visualization (equal). Alejandro Strachan: Conceptualization (equal); Funding acquisition (equal); Project administration (equal); Resources (equal); Supervision (equal); Writing – review \& editing (equal). Kenneth H. Sandhage: Conceptualization (equal); Funding acquisition (equal); Project administration (equal); Resources (equal); Supervision (equal); Writing – review \& editing (equal). Michael S. Titus: Conceptualization (equal); Funding acquisition (equal); Project administration (equal); Resources (equal); Supervision (equal); Writing – review \& editing (equal).

\section*{Conflicts of interest}
There are no conflicts to declare.

\section*{Data availability}
The code used for generating descriptors, the training dataset, and the predicted new alloy dataset with their corresponding descriptors, which support the findings of this study, are openly available in a public repository at \url{https://github.itap.purdue.edu/michaeltitusgroup/pyTCPlotter} under the ML-Descriptors folder.

\section*{Acknowledgements}

We acknowledge support from the U.S. National Science Foundation, DMREF program, under Contract No. 1922316-DMR.

\bibliographystyle{unsrt}
\bibliography{references-2}

% --- START OF APPENDIX SECTION ---

\appendix 
\newpage

\section{Supplementary Information}

\renewcommand{\figurename}{SI Figure}
\renewcommand{\thefigure}{\arabic{figure}}
\renewcommand{\tablename}{SI Table}
\setcounter{table}{0}
\setcounter{figure}{0} % or 0, depending on where you start
\begin{figure}[H]
  \centering
  \includegraphics[width=\textwidth]{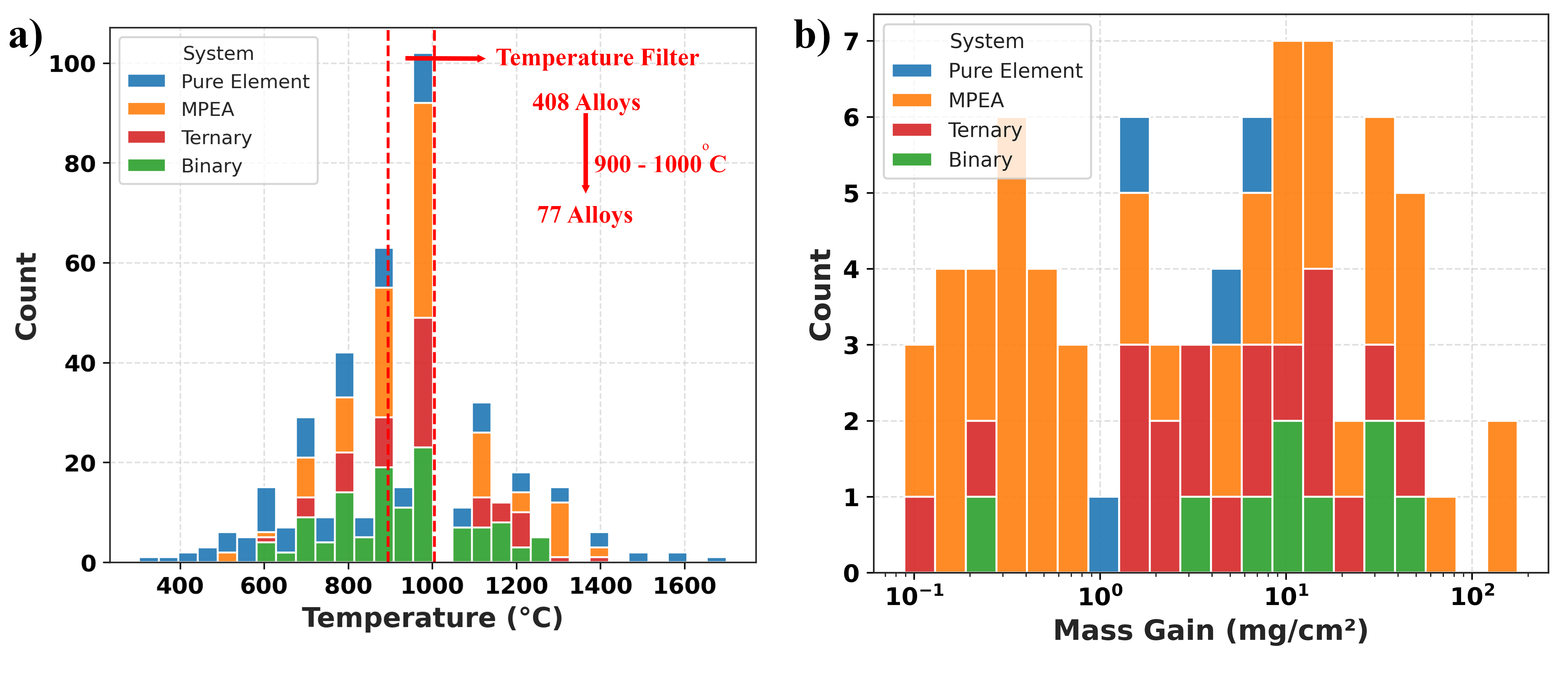}
  \caption{(a) Bar chart showing the distribution of different alloy systems and oxidation temperatures present in the entire dataset. (b) Bar chart showing the distribution of specific mass gain values for the selected 77 alloys oxidized at 900–1000°C} 
  \label{fig:data}
\end{figure}
\begin{figure}[H]
  \centering
  \includegraphics[width=\textwidth]{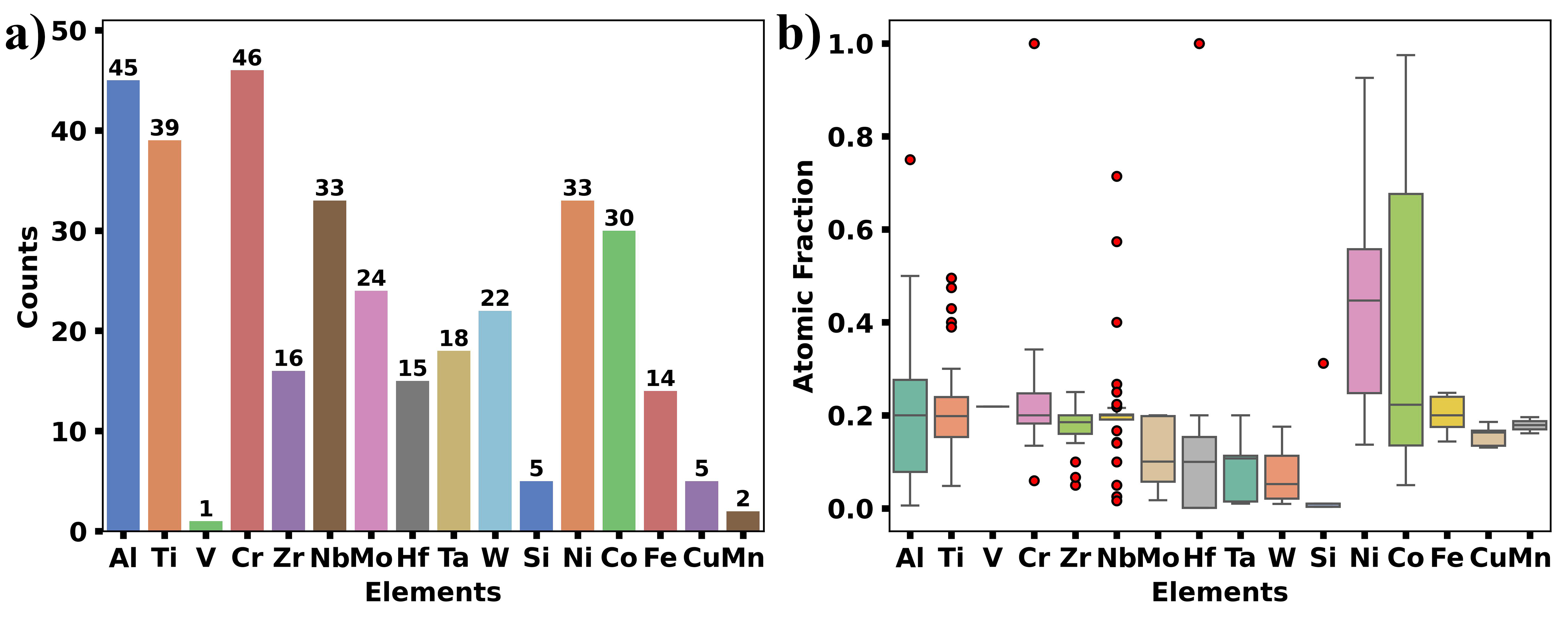}
  \caption{(a) Bar chart showing the number of alloys (out of a total of 77 alloys) containing a given element (present at any concentration) in the dataset that had been filtered for oxidation at 900–1000°C. (b) Box plot showing the atomic fraction distributions of elements in this dataset.}
  \label{fig:elements}
\end{figure}
\begin{table}[H]
    \renewcommand{\arraystretch}{1.8} % Adjust the row spacing
    \centering
    \caption{Metal alloy-based descriptors used in this study.}
    
    \resizebox{\textwidth}{!}{ % Rescale the table to fit within the page width
    \begin{tabular}{@{\hskip 5pt}c@{\hskip 5pt}c@{\hskip 5pt}>{\centering\arraybackslash}p{8cm}@{}}
        \toprule
        \textbf{Notation} & \textbf{Formalism} & \textbf{Description} \\ 
        \midrule
        $T_m$            & $\sum_{i=1}^n c_i T_{m,i}$ & Average melting temperature \\ 
        $V_{\text{misfit}}$ & $\sum_{i=1}^{n} (\bar{V} - V_i)^2$ & Atomic volume misfit \\ 
        $\overline{R}$    & $\sum_{i=1}^n c_i r_{at,i}$ & Average atomic radius\\ 
        $\delta$          & $\sqrt{\sum_{i=1}^n c_i \left( 1 - \frac{r_i}{\overline{r}} \right)^2} \times 100$ & Asymmetry of atomic radius\\ 
        $VEC$             & $\sum_{i=1}^n c_i VEC_i$ & Average valence electronic concentration \\ 
        $\Delta S_{\text{mix}}$ & $-\sum_{i=1}^n c_i \ln c_i$ & Entropy of mixing\\ 
        $|Y|$             & $\sqrt{\sum_{i=1}^n c_i \left( 1 - \frac{Y_i}{\overline{Y}} \right)^2}$ & Asymmetry of Young's moduli. \\ 
        \midrule
        $\Delta Y$        & $Y_{\text{i,max}} - Y_{\text{i,min}}$ & Range of Young's moduli \\ 
        $\Delta \rho$     & $\rho_{\text{i,max}} - \rho_{\text{i,min}}$ & Range of density \\ 
        $\Delta T_m$      & $T_{m,\text{i,max}} - T_{m,\text{i,min}}$ & Range of melting temperature \\ 
        $\Delta r_{at}$   & $r_{at,\text{i,max}} - r_{at,\text{i,min}}$ & Range of atomic radii \\ 
        $\Delta K$        & $K_{\text{i,max}} - K_{\text{i,min}}$ & Range of bulk moduli \\ 
        $\Delta VEC$      & $VEC_{\text{i,max}} - VEC_{\text{i,min}}$ & Range of VEC \\ 
        \midrule
        $\rho$            & -- & Bulk density of the alloy, kg/m\(^3\)\\ 
        \textit{Solidus}  & -- & Solidus temperature of the alloy \\ 
        \textit{Liquidus} & -- & Liquidus temperature of the alloy \\ 
        $\Phi$ & - & Reduced phase one-hot-encoding\\
        %$[1, 0, 0, 0, 0, 0, 0, 0]/[0, 1, 0, 0, 0, 0, 0, 0] etc.$ & -- & Reduced Phase One-Hot-Encoding (O.H.E.)\\
      
        \bottomrule
    \end{tabular}
    }
    
    \label{tab:descriptors}
\end{table}
\begin{figure}[H]
\centering
\includegraphics[width=\textwidth]{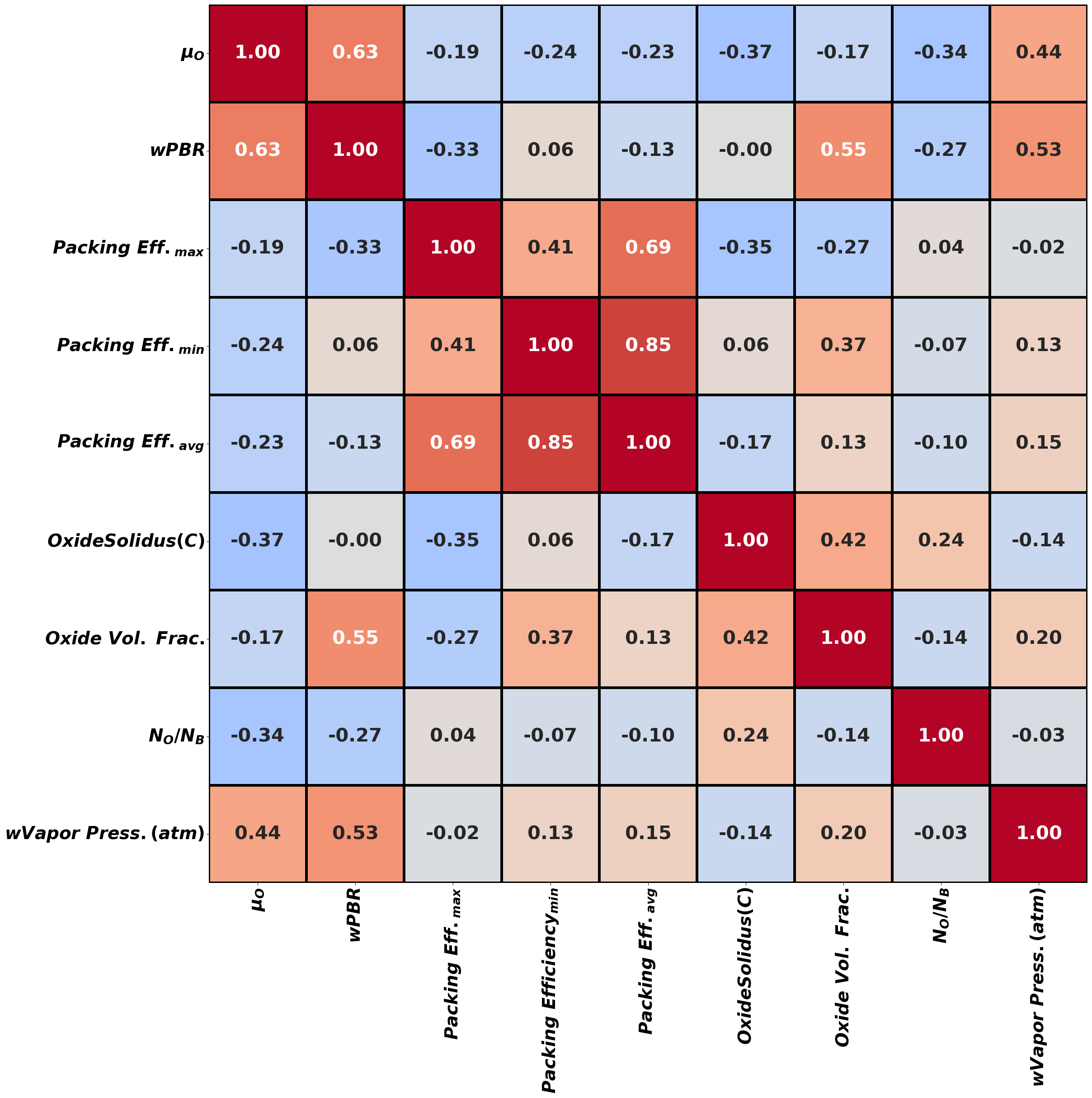}
\caption{
 Pearson correlation heatmap of oxidation-related descriptors(for the training data) used in the Gaussian Process Regression (GPR) model. The matrix quantifies pairwise linear correlations between variables such as oxygen chemical potential ($\mu_\mathrm{O}$), weighted Pilling–Bedworth ratio (wPBR), atomic packing efficiency (max, min, and avg), solidus temperature of the oxide mixture (°C) in the \FTH{}, total volume fraction of the oxides int \FTH{}, oxygen solubility ratio $N_\mathrm{O}/N_\mathrm{B}$, and weighted vapor pressure (atm). Strong positive correlations are observed among the packing efficiency metrics (e.g., $\rho = 0.85$ between min and avg), while $\mu_{O}$ and wPBR exhibit moderate correlation ($\rho = 0.63$). In contrast, oxide solidus temperature and $\mu_\mathrm{O}$ show weak to moderate negative correlation, indicating potential trade-offs in design targets for oxidation resistance.
}
\end{figure}

\begin{figure}[H]
\centering
\includegraphics[width=\textwidth]{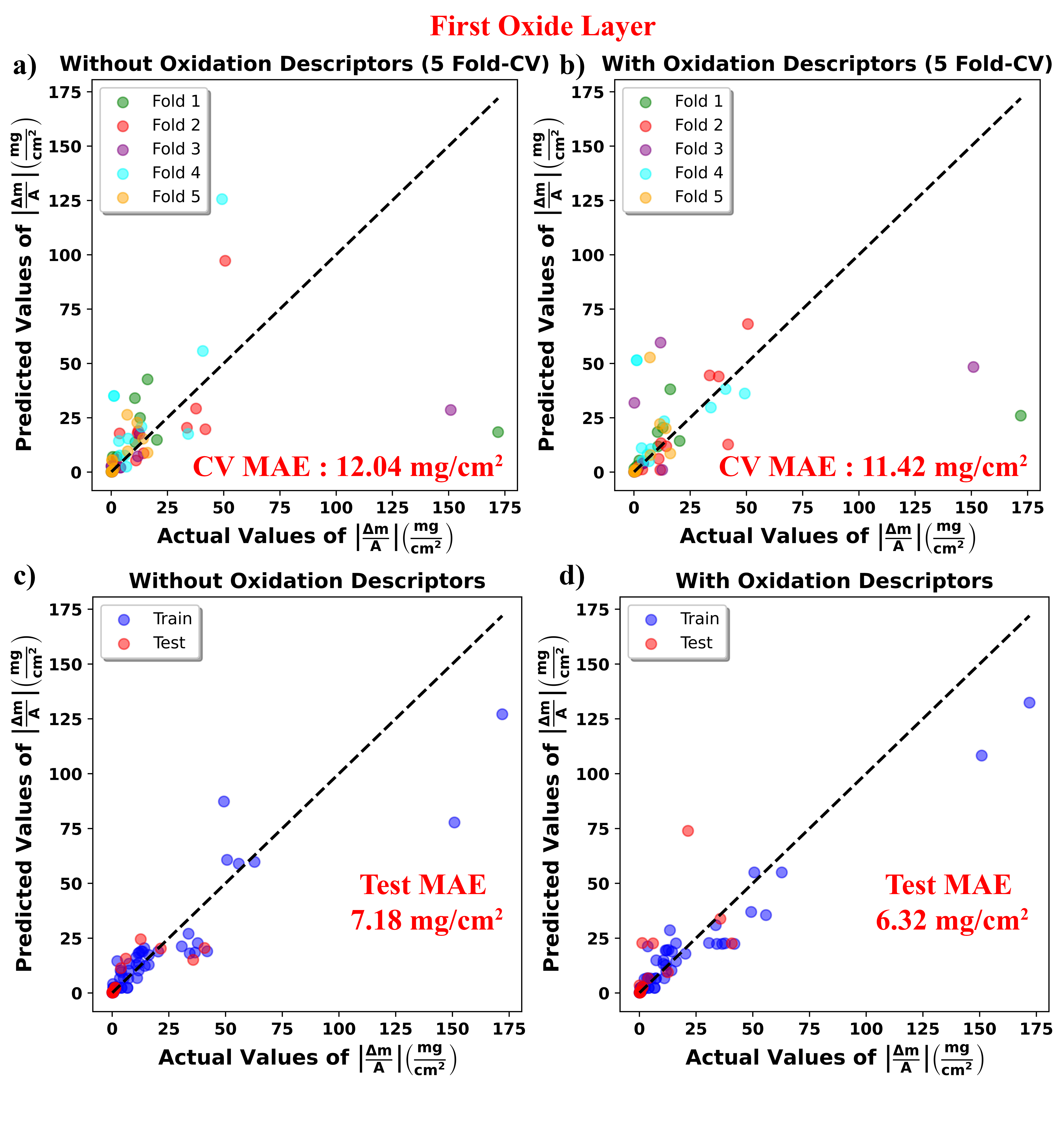}
\caption{
 Parity plots comparing the predictive performance of Gaussian Process Regression (GPR) models for specific mass gain of the \textbf{first oxide layer} using input features with and without oxidation-related descriptors. sub-figures~(a) and~(b) show 5-fold cross-validation results with CV MAEs of 12.04 mg/cm\(^2\) and 11.42 mg/cm\(^2\), respectively. sub-figures~(c) and~(d) depict the model performance on the combined train, and test datasets, where the inclusion of oxidation descriptors improves the Test MAE from 7.18 mg/cm\(^2\) to 6.32 mg/cm\(^2\). While improvements are observed in both CV and test scenarios, the performance trends mirror those of the \FTH{} model in Figure~\ref{fig:GPR}, reaffirming the modest but consistent benefit of oxidation descriptors. However, predictions remain concentrated in the low mass-gain regime, indicating similar challenges in extrapolation. This suggests that despite the layer-specific modeling approach, improvements in predictive accuracy still depend on the diversity and balance of training data.
}
\end{figure}

\end{document}